# Any DOF all at once: single photon state tomography in a single measurement setup

ROEY SHAFRAN,[1,†] RON ZIV,[1,†] MORDECHAI SEGEV[1,2,*]

[1]*Department of Electrical and Computer Engineering, Technion – Israel Institute of Technology, Haifa 32000, Israel*
[2]*Physics Department, Technion – Israel Institute of Technology, Haifa 32000, Israel*
[†]*These authors contributed equally to this work*
**msegev.technion.ac.il*

**Abstract:** Photonic quantum technologies utilize various degrees of freedom (DOFs) of light, such as polarization, frequency, and spatial modes, to encode quantum information. In the effort of further improving channel capacity of quantum communication, and for increasing the complexity of available quantum operations, high-dimensional and hyperentangled states are now gaining interest. However, efficiently measuring these high dimensional states is challenging due to the large number of measurements required for reconstructing the full density matrix via quantum state tomography (QST), and the fact that each measurement requires some modification in the experimental setup. Here, we propose a framework for reconstructing the density matrix of a single-photon hyperentangled across multiple DOFs using a single intensity-measurement obtainable from traditional cameras, and discuss extensions for multiphoton hyperentangled states. Our method hinges on the spatial DOF of the photon and uses it to encode the quantum information from the other DOFs. We numerically demonstrate this method for single-photon OAM-spin and OAM-frequency entangled states using an ideal coupler and a multimode fiber, to perform the information mixing and transfer the encoding to spatial information, where it is detected using a simple camera. This technique simplifies the experimental setup and reduces acquisition time compared to traditional QST-based methods. Moreover, it allows the recovery of DOFs that conventional cameras cannot detect, such as polarization, thus eliminating the need for projection measurements.

## 1. Introduction

Photonic quantum technologies have gained significant interest in recent years due to the favorable properties of photons, among them lack of decoherence, scalability and compatibility with standard technologies such as optical fibers and integrated photonic circuits [1,2]. Therefore, there is clear interest in developing photonic-based applications in many areas of quantum technology ranging from quantum communication [3] to simulations of complex quantum systems [4] and quantum computation [5] .

At the base of quantum information technologies stands entanglement, which relies on the physical degrees of freedom (DOFs) in which quantum information (e.g., qubits) can be encoded. In photonics, commonly used DOFs include polarization and momentum, the latter offering spatial path entanglement [6,7]. Yet the complex structure of light offers many additional DOFs, all controllable, and each offering a Hilbert space suitable for quantum information encoding. Examples include frequency and time-bins (using temporal and spectral structures) [8,9], and spatial DOFs such as orbital angular momentum (OAM) [10–13]. Moreover, to further increase the information density per physical constituent of the system, it is possible to use DOFs that offer more than two levels per particle, for example OAM, and move the information coding from qubits to high dimensional qudits, or to utilize states with multiple DOFs per particle. The latter, known as hyperentangled states [14], involve entanglement across multiple DOFs, have been gaining much interest recently. The use of hyperentanglement open the door to increased performance in virtually all applications of quantum information, ranging from higher channel capacity [15,12,16–19] and enhanced security [20] to noise robustness [13,21] in quantum communication, and allows for more complex [22,8,16] and deterministic quantum operations [8,23,24].

Naturally, the use of entangled states requires means to measure them in an efficient manner and recovering the information they encode. Standard approaches for reconstructing photonic states rely on quantum state tomography (QST) [25–28], which requires measurements of the complete set of observables accounting for all the real values of the density matrix representing the quantum state. In an experimental setup, probing different observables amounts to varying the configurations of polarizers and spatial light modulators for different spin-spatial modes [27,28], or phase shifters and interferometers [8,9] in the case of frequency/time bins, and click detectors to record coincidences in the case of multi-photon states. This poses a problem even when considering the relatively simple case of qubits. Namely, recovering the density matrix of $n$ qubits requires $2^{2n}$ measurements, which becomes exceedingly difficult for large $n$. Using high-dimensional DOFs and hyperentanglement instead of qubits makes this problem far more severe even for small $n$, as the number of required measurements scales linearly with the dimensionality of each DOF and exponentially with the number of entangled DOFs. For instance, characterizing a photon entangled in polarization and a 2-level OAM system requires 16 measurements [29]. If the OAM system is extended to 6 levels of OAM, the number of measurements increases to 144. Introducing an additional DOF, such as a 4-level frequency-bin system, raises the requirement to 256 measurements. Adding more DOFs and increasing each of their dimensions leads to more measurements, making the tomography process unfeasible. Obviously, having to carry out so many different measurements in different configurations also induces noise into the system, while also increasing the calibration needs of the setup. Clearly, it is highly desirable to reduce the number of measurements required to fully characterize a quantum state, especially when it is hyperentangled.

These insights motivated a rich body of work aiming at simplifying the measurement setup and reducing the number of different quantities needed to characterize the quantum state. In a broad scope, modern scalable theoretical frameworks have been proposed as tools to optimize the total amount of data required for state reconstruction. For example, compressed sensing tomography uses prior information on the state's rank [30,31], while adaptive tomography [32,33] determines the next measurement to perform based on the results of previous measurements. In a different manner, threshold QST [34,35] allows the user to balance between reconstruction accuracy and the number of measurements, prioritizing the most significant ones, and if only a function of the state is needed, full reconstruction can be avoided by using shadow tomography [36–38]. In the realm of photonics, previous works achieved reconstruction using a single measurement setup by leveraging known state structures [39] and ancilla modes [40,41], and exploited spatial structure to enable tomography with minimal projections or reference interference [42,43]. However, the recovery of a quantum state displaying entanglement between multiple DOFs using a single measurement setup has never been proposed.

**Here, we propose a general framework for reconstructing the density matrix of a single-photon entangled across multiple DOFs, using a single intensity image obtained with a single experimental setup.** Additionally, we provide extensions for multiphoton hyperentangled states by transitioning from intensity to coincidence measurements. Our framework utilizes the inherent parallelism of intensity measurements provided by a camera – where a single image represents a set of simultaneous projections onto multiple spatial modes. We leverage the spatial DOF of the photon to encode the information transferred to the spatial structure of the photonic wavefunction from the other DOFs, and utilize the coupling between the different spatial DOFs for complete recovery of the state. We demonstrate our method numerically for single-photon qudits entangled in OAM-spin and in OAM-frequency-bin, showcasing results for both an ideal random coupler and a physical realization based on a multi-mode fiber. Necessitating only a single fixed experimental setup, our framework can significantly simplify experimental settings and reduce acquisition time, compared to traditional click-based or projective measurement schemes. Remarkably, our method enables the recovery of DOFs for which standard imaging devices are blind to, such as polarization, exemplifying the ability to remove the need for projection measurements for the complete recovery of quantum information borne on single photons.

## 2. Theory

In this section, we develop the mathematical framework for state reconstruction using spatial intensity measurements. We begin by defining the relationship between spatial modes and the probability distribution measured at an imaging plane, namely the measured intensity at a given pixel $(x, y)$, informally given as $Tr[|x, y\rangle\langle x, y|\rho]$, and will be formally defined later. Subsequently, we analyze why direct intensity measurements often fail to be informationally complete (IC), and introduce the concept of a coupling system to overcome this limitation. Finally, we formulate the reconstruction as a constrained optimization problem. A conceptual overview of this protocol – from state propagation through the coupler to the final density matrix estimation – is illustrated in Figure 2.

### 2.1. Spatial Modes and Intensity Measurements

We begin by defining the spatial photon modes, following the formulation in [44], which dictate the probability distribution of a photon at some plane in free space (henceforth denoted as the "imaging plane"). A single-photon excitation in mode $m$ is represented by $|f_m\rangle = \hat{a}_m^\dagger|0\rangle$, where $\hat{a}_m^\dagger$ and $\hat{a}_m$ are the creation and annihilation operators and the function $f_m$ represents a normalized solution to the Maxwell equations. In many optical systems, particularly those involving imaging or free-space propagation, it is natural to consider spatial modes where each mode function $f(r_\perp)$ depends only on the spatial coordinates transverse to the main propagation direction. A common basis for such systems is the Laguerre Gauss (LG) family of spatial functions.

The observable associated with measuring a photon in position $r_\perp$ is given by

$$\hat{N}(r_\perp) = \sum_{m,m'} \hat{a}_m^\dagger \hat{a}_{m'} f_m^*(r_\perp) f_{m'}(r_\perp) \tag{1}$$

where $\{f_m(r_\perp)\}_{m=1}^{d}$ spans a d-dimensional spatial Hilbert space $\mathcal{H}_{spatial}^{(d)}$. For a given state $\rho$, the probability of detecting a photon at position $r_\perp = (x, y)$ is expressed as:

$$\mathrm{P}(r_\perp) = Tr\left(\rho\widehat{N}(r_\perp)\right) = Tr(\rho\, |r_\perp; d\rangle\langle r_\perp; d|) \quad (2)$$

Here, we introduce the pixel state $|r_\perp; d\rangle = \sum_{l=1}^{d} f_l^*(r_\perp)|f_l\rangle$, which represents the superposition of spatial states at a specific point. Given a pixel with area $\mathcal{S}$, the probability of detecting a photon is $\int_{\mathcal{S}} P(r_\perp)dr_\perp = Tr\left(\rho \int_{\mathcal{S}}|r_\perp; d\rangle\langle r_\perp; d|dr_\perp\right)$. For simplicity, we assume a sufficiently small $\mathcal{S}$ such that $\int_{\mathcal{S}} P(r_\perp)dr_\perp \approx P(r_\perp)$. However, this assumption is not essential, and the following theory remains valid without it.

In a physical setup, a camera image is formed by $N$ independent draws from this distribution across $n$ pixels at positions $\{r_i\}_{i=1}^{n}$. Of course, since any quantum measurement is an expectation value (the mean over multiple incidents), in the case of a camera or multiple detectors – the expectation value is the mean over sufficient incidents (clicks) on each pixel. By defining the positive semidefinite operators $\Pi_i = |r_\perp; d\rangle\langle r_\perp; d|$, this measurement scheme corresponds to a Positive Operator-Valued Measure (POVM) [45], provided the image area covers the support of all modes such that $\sum_{i=1}^{n} P(r_i) = 1$, and $\sum_{i=1}^{n} \Pi_i = I$.

From Equation (2) we observe that an intensity image obtained from a camera in a single experimental setup corresponds to a set of parallelized projection measurements performed simultaneously.

## 2.2. The need for a coupler: intensity measurements are not IC

For successful state reconstruction, the POVM must be informationally complete (IC), ensuring that no two distinct density matrices result in the same spatial intensity distribution [46]. If two states are different, there exists some physical observable that can distinguish between them. Intuitively, an IC POVM must be able to represent every such observable. Mathematically, this translates to the set $\{\Pi_i\}_{i=1}^{n}$ spanning the space of Hermitian operators acting on $\mathcal{H}_{spatial}^{(d)}$ [47–49]. Weaker conditions can be found if some prior information on the state exists, such as a bound on its rank [49,50].

Direct intensity measurement alone is insufficient if the spatial modes do not satisfy the IC condition. A typical example of such a fail-case is found in paraxial OAM states, defined as:

$$LG_{lp} \propto \exp\left[-\frac{r^2}{w^2(z)}\right] L_p^{|l|}\left(\frac{2r^2}{w^2(z)}\right)\exp[il\phi] \quad (3)$$

where $w(z) = w(0)\sqrt{(z^2 + z_R^2)/z_R^2}$ with $w(0)$ the beam waist, $z_R$ the Rayleigh range, and $L_p^{|l|}$ is an associated Laguerre polynomial [51]. Because the sign of $l$ only affects the phase, two OAM modes with opposite momentum possess identical intensity profiles, meaning the POVM they define is not IC (see Fig. 1(a)). For instance, in [52] the reconstruction of spatially structured light required a second intensity measurement, applying a unitary transformation $\mathcal{U}$ on the state beforehand. While this doubled the number of measurements, it formed an IC set allowing full recovery.

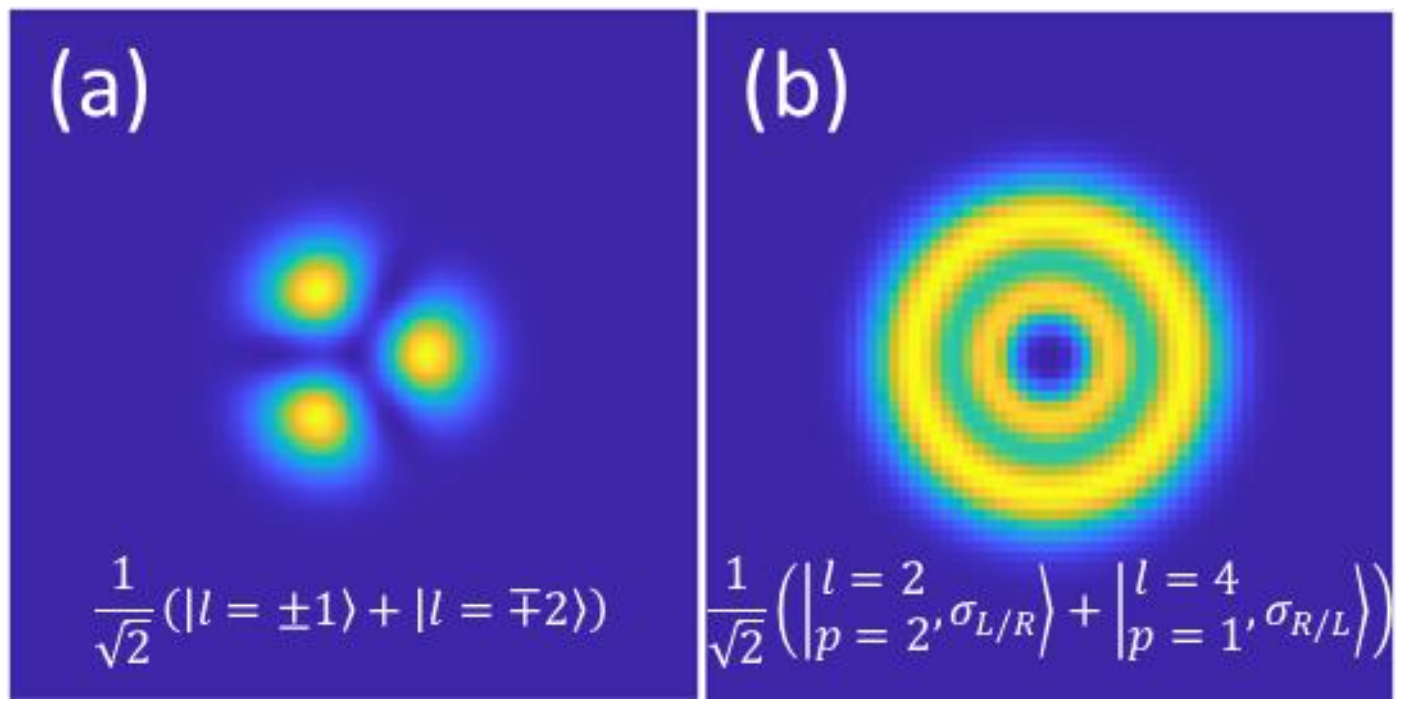

**Fig. 1**. Examples of fail cases for distinguishing states from intensity measurements. (a) Intensity ambiguity when the spatial modes do not hold the condition for IC. The OAM states $|\pm1\rangle$ and $|\pm2\rangle$ have the same intensity profile resulting in the two states $1/\sqrt{2}\,(|l = \pm1\rangle + |l = \mp2\rangle)$ producing the same measurement. (b) The two hyperentangled states only differ in their spin DOF, resulting in the same intensity measurement even though the spatial modes $\{|l = 2, p = 2\rangle, |l = 4, p = 1\rangle\}$ hold the condition for IC separately.

The challenge is more severe for hyperentangled states. Consider a state $\rho$ prepared in the composite Hilbert space $\mathcal{H}_{spatial}^{d} \otimes \mathcal{H}_{non-spatial}^{m}$, where m represents non-spatial DOFs like frequency, polarization, time bins etc. Importantly, $\mathcal{H}_{non-spatial}^{m}$ can be a tensor product of multiple non-spatial DOFs spaces. Because standard cameras are blind to non-spatial DOFs, the measured intensity is determined only by the partial trace over the non-spatial Hilbert space:

$$P(r_\perp) = Tr(Tr_m(\rho)\,|r_\perp; d\rangle\langle r_\perp; d|) \tag{4}$$

Thus, direct measurement can only recover $Tr_m(\rho)$, leaving the correlations between spatial and non-spatial DOFs hidden. To address this, we leverage the (theoretically) infinite space of spatial modes. We can consider $\rho$ as a state in a larger Hilbert space $\mathcal{H}^D_{spatial} \otimes \mathcal{H}^m_{non-spatial}$, introducing additional $D - d > 0$ ancilla modes – where their corresponding entries in $\rho$ are zero. While the natural spatial structure of light offers us these ancilla modes, to be of use for reconstruction – the information from non-spatial DOFs must be mapped onto these ancillas.

### 2.3. Coupler lifting and reconstruction formulation

To recover the hidden information, we propose propagating the state through a coupling system $\mathcal{U}$ that mixes spatial and non-spatial DOFs. This results in a modified spatial distribution:

$$P(r_\perp) = Tr(Tr_m(\mathcal{U}\rho\,\mathcal{U}^\dagger)\,|r_\perp; D\rangle\langle r_\perp; D|) \tag{5}$$

Recovering a $d \times m$ dimensional state by encoding it into a D-dimensional state requires the necessary dimensionality condition:

$$D \geq dm \tag{6}$$

The coupled measurement can be presented as a POVM (see Appendix A):

$$\Pi_i(\mathcal{U}) = \mathcal{U}^\dagger(|r_\perp; D\rangle\langle r_\perp; D| \otimes I_m)\,\mathcal{U} \tag{7}$$

where $I_m$ is the identity operator in the non-spatial DOF space. The probability of a photon incident on pixel at position $r_i$ is then given by:

$$P(r_i) = Tr\big(\rho\,\Pi_\mathrm{i}(\mathcal{U})\big) \tag{8}$$

Crucially, while (6) is a necessary condition, full reconstruction is guaranteed if and only if $\{\Pi_i(\mathcal{U})\}_{i=1}^n$ is IC, spanning the space of Hermitian operators of dimension $dm \times dm$. Formulating the process as a POVM (Eq. (7)), rather than the representation in Eq. (5), is not only mathematically elegant but also significantly reduces computational cost, by allowing to precompute the measurement operators (Appendix B).

Our reconstruction protocol (Fig. 2) estimates the distribution from an intensity image $\{I(r_i)\}_{i=1}^n$ as $P(r_i) = I(r_i)/N$. Note that, even if in practice the number of photons is not known precisely, one can approximate it by normalizing the total signal obtained on the camera over the entire integration time, yielding a normalized intensity representing the probability. We formalize the recovery as a constrained least-squares problem:

$$\min_{\rho} \sum_{i=1}^{n} \left| Tr\big(\rho\,\Pi_i(\mathcal{U})\big) - \frac{I(r_i)}{N} \right|^2 \tag{9}$$
$$s.t \quad \rho = \rho^\dagger, \quad \rho \geq 0, \quad Tr(\rho) = 1$$

This formulation is standard and can be solved by a variety of methods and approaches. Henceforth, we use CVX [53,54], a package for specifying and solving convex programs.

## 3. Results

In this section, we present the reconstruction performance of our method using simulated data. We evaluate the framework across different coupling systems and photonic DOFs to demonstrate its generalizability, specifically focusing on states entangled across both spatial and polarization (spin) DOFs. We begin with a general case using an ideal random coupler. Subsequently, we extend the methodology to a physically realizable system utilizing the experimentally measured transmission matrix (TM) of a multi-mode fiber (MMF). For completeness, Appendix C details how the fiber medium can be extended, via cross-phase modulation, to recover the frequency DOF of light.

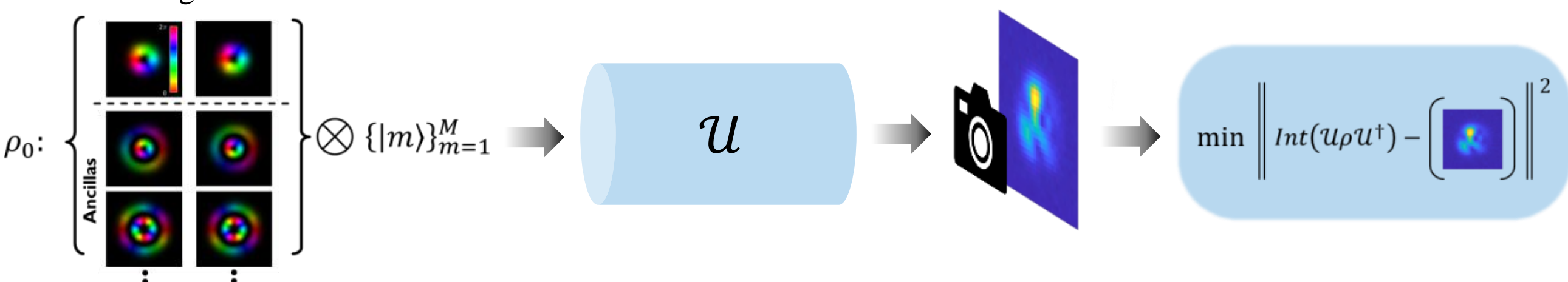

**Fig. 2.** Reconstruction scheme. A state propagates through a coupling system mixing its information onto the spatial modes and afterwards it is imaged onto a camera. The state is recovered by solving the optimization problem minimizing the L2-error between the measured and estimated intensity image.

### 3.1. Random Coupler

We first evaluate our method under general conditions using a simulated random coupler uniformly sampled from the unitary group [40]. This ensures that the information encoded in the quantum state is well-distributed across the output spatial modes, resulting in POVM elements that are as uncorrelated as possible. For these simulations, we utilize LG modes with a constant total order $2p + |l|$, which form an IC set [52]. We note, however, that choosing an IC set is not strictly necessary. A non-IC set of spatial modes simply requires the coupler to have more ancilla modes for full reconstruction.

To analyze performance, we simulate a quantum state $\rho_0$ of dimension $d \times m$ (where $d$ represents spatial modes and m non-spatial modes) of both rank 1 (pure) and full-rank. The state is passed through a coupler that mixes its information onto $D$ spatial modes and a $32 \times 32$ intensity image is formed by $n$ detected photons. To simulate realistic conditions, additive white Gaussian noise is included (e.g., SNR of 30 dB) as shown in Figure 3(a). The reconstruction can be seen visually in Figure 3(b) where the state was recovered with 0.97 fidelity.

The reconstruction performance is quantified by the fidelity, $F(\rho_0, \rho_{rec}) = \left( Tr\sqrt{\rho_0^{\frac{1}{2}} \rho_{rec} \rho_0^{\frac{1}{2}}} \right)^2$, with each data point representing the mean fidelity, averaging over 50 random state initializations. As shown in Fig. 3(c), for $10^5$ photons, all states are recovered with a mean fidelity of over 0.98 regardless of rank. This supports the intuition that increasing the number of ancilla modes provides measurement redundancy, making the estimation more robust to errors. In Figure 3(d), we see that the fidelity increases sharply and then plateaus with the number of additional ancilla modes as they provide the redundancy needed to maintain high fidelity in the presence of noise.

### 3.2. OAM-Spin States

We next consider states entangled in their spin and OAM DOFs, which was recently shown experimentally using metamaterials [29] to encode two qubits on a single photon. To recover a state $\rho \in \mathcal{H}_{OAM}^{d} \otimes \mathcal{H}_{spin}^{2}$, we utilize the intrinsic spatial and polarization mode coupling of a MMF [55]. The classically acquired transmission matrix (TM) of a multimode fiber has been shown to allow control of quantum light inside the fiber [56–58]. Thus, to model our forward process in Equation (7) we use the experimentally measured TM of a 2-meter long, 306-mode graded-index fiber [55] as our unitary $\mathcal{U}$.

The practical realization of this scheme is well-supported by current technology. Modern sensors, such as Electron-Multiplying Charge-Coupled Device (EMCCD) cameras, provide the high pixel density and single-photon sensitivity required for accurately imaging the spatial structure of single photons [59–62]. A critical requirement, is that the coupler must "mix" the states well enough to satisfy the condition for an IC POVM.

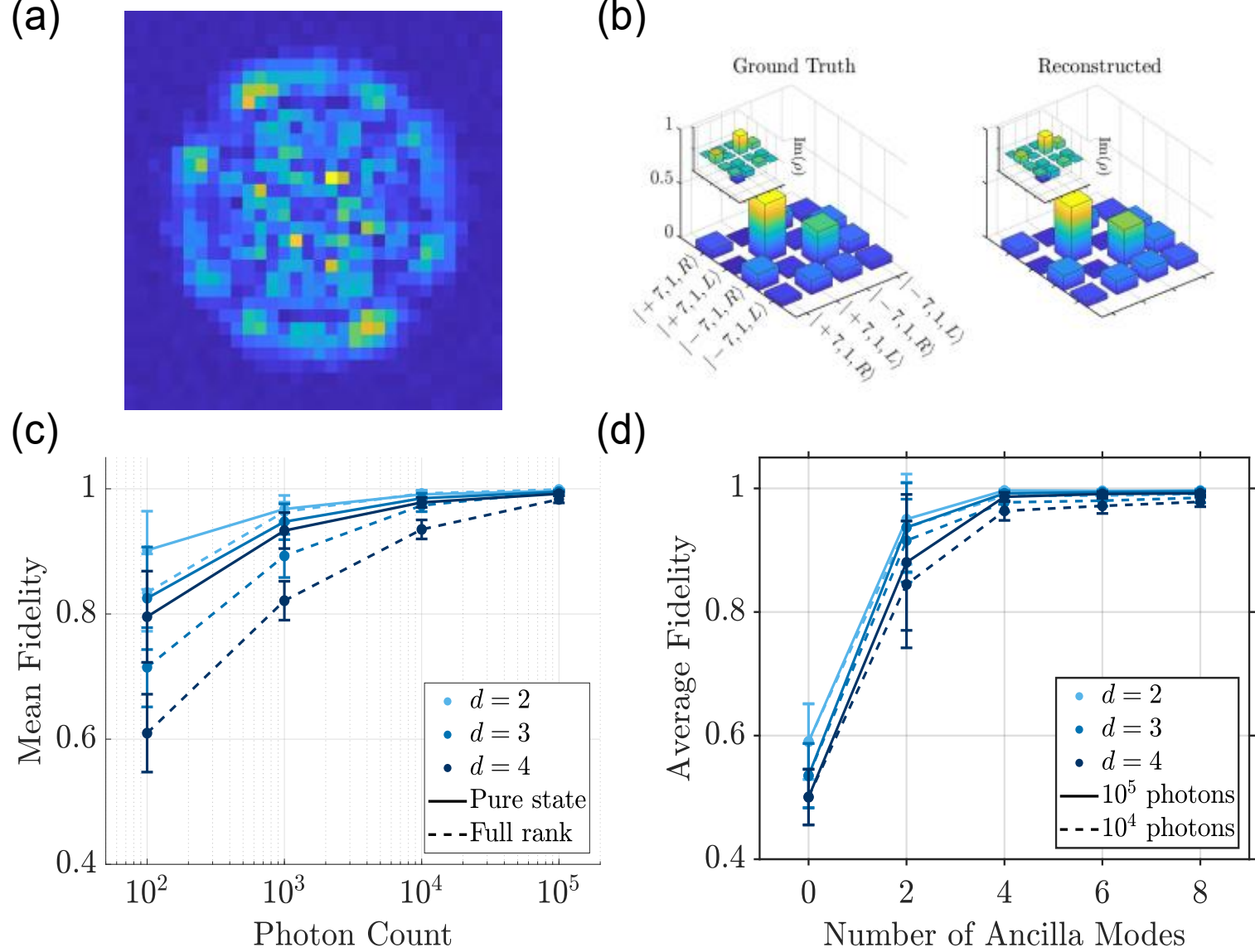

Figure 3: **Performance analysis using a random unitary coupler.** (a) Intensity image of a 4-dimensional OAM-spin state ($d = 2, m = 2$) with $D = 8$ spatial modes (6 ancilla modes), $10^4$ photons, and 30dB SNR. (b) Comparison of ground truth (left) and reconstructed (right) density matrices (Fidelity = 0.97) for the state in (a). (c) Mean fidelity bs. photon count for dimensions $d = 2,3,4$ and $m = 2$, comparing pure (solid) and full-rank (dashed) states using a coupler of 8 ancillas. (d) Mean fidelity vs. additional ancilla modes ($D - d$) for rank-1 states at $10^4$ and $10^5$ photons. The performance stabilizes as the number of ancillas grows, providing the necessary redundancy for robust recovery.

Notably, we have verified that the experimentally measured TM used in our simulations satisfies this condition, as the resulting POVM operators span the space of Hermitian operators for the reconstructed dimensions and spatial modes.

In an ideal graded-index fiber, OAM modes are organized into mode groups with low inter-group crosstalk [63], where the modes in each group share a constant total order defined by $|l| + 2p$. For the 306-mode fiber used in our simulations, the modes are divided into 17 mode groups, where the number of spatial modes in the $p^{th}$ group is exactly $p$. The block-diagonal structure of the TM in Fig. 4(b) visually confirms this mode-group isolation. Because of the low crosstalk, the effective ancilla space available for a given state is largely determined by the number of modes within its constituent mode group. Looking at the reconstruction results in Fig. 4(a-b), we observe that this modal structure of the fiber is fundamentally linked to the reconstruction fidelity. States utilizing OAM modes from lower-order mode groups (e.g., $l = \pm 2, p = 1$) have access to a smaller pool of ancilla modes, resulting in a lower redundancy that makes them more sensitive to noise. In contrast, states coupled to higher-order mode groups benefit from a larger effective number of ancillas as they contain a larger number of modes, leading to higher fidelity at lower photon counts. This relationship is highlighted by the contrast between $d = 2$ states (Fig. 4a) and $d = 4$ states (Fig. 4b). In the latter case, mode groups that were previously sufficient for $d = 2$ states (e.g., Group 6) now fail to reach unity fidelity (represented by dashed lines) as they violate the dimensionality condition ($D < dm$). This physical insight suggests that the mixing properties of the system could be further enhanced by introducing intentional perturbations to the fiber, such as bending or increasing ellipticity, to increase crosstalk between different mode groups [63], effectively increasing the set of ancillas for any given state.

Another consideration demanding a discussion is that the accuracy of our reconstruction relies on the precision of the characterization of $\mathcal{U}$ and the temp oral stability of the experimental setup. For reliable tomography, the system must remain stable during the measurement. MMFs are particularly suitable for this framework as their TMs can be accurately characterized and remain stable over relevant experimental timescales [57,58]. To highlight the relevance of MMFs to our proposed methodology, we analyze and quantify the impact of experimental inaccuracies by simulating deviations between the actual fiber transformation and the one used in the reconstruction algorithm. As detailed in Appendix C, our analysis demonstrated that the methodology maintains high fidelity across a realistic range of characterization errors.

Having demonstrated reconstruction of OAM-spin states with a multi-mode fiber coupler and analyzed considerations for its practical realization, we note that our framework can be extended to frequency DOFs as well. A detailed discussion and numerical simulations, including spatio-spectral coupling via Kerr-induced cross-phase modulation, are provided in Appendix C. Importantly, this implies that this approach could be extended to recover states entangled across all three DOFs - spatial, spectral, and spin.

## 4. Multiphoton Hyperentangled States

Finally, we extend our analysis to the multiphoton regime, examining the limitations and requirements for tomography of hyperentangled states with a single measurement setup. For clarity, we consider a bi-photon state, of photons A and B, in the cross-product space $\mathcal{H}_A \otimes \mathcal{H}_B$ of two identical spaces of $d$ spatial and $m$ non-spatial modes in each. What information is obtained by naively using the single photon scheme on this state? Passing both photons through the coupling system, the intensity at each pixel no longer corresponds to a standard POVM.

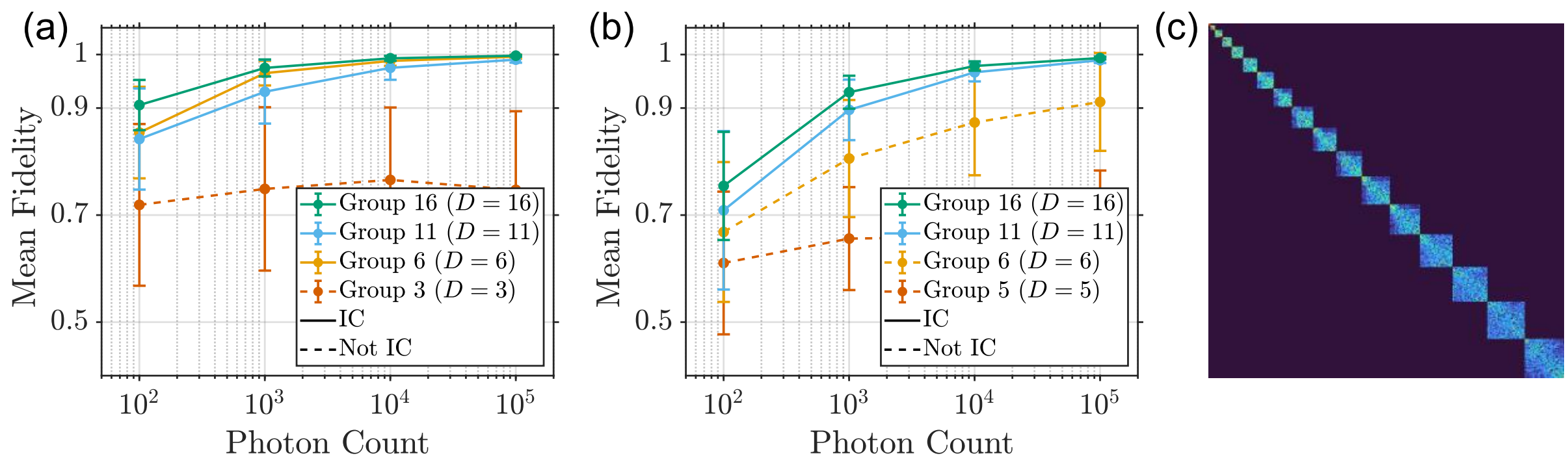

Figure 4: **OAM-Spin state reconstruction in a graded-index multimode fiber (MMF)**. Mean fidelity vs photon count for (a) 4 dimensional and (b) 8-dimensional hyperentangled states across different mode groups, with higher-order groups showing increased robustness due to a larger effective ancilla space. Solid lines denote IC groups ($D \geq dm$), while dashed lines indicate non-IC groups. (c) Absolute value of the 306-mode experimental transmission matrix. The block-diagonal pattern highlights the 17 isolated mode groups of the fiber.

This is because we cannot distinguish whether the two photons are incident on a given pixel or if two photons from two different pairs of photons incident at the same pixel. Respectively, the probabilities of a single photon hitting pixel $i$ and of both photons hitting it are:

$$P_1 = Tr\left(\frac{1}{2}\left(\Pi_i(\mathcal{U}) \otimes I + I \otimes \Pi_i(\mathcal{U})\right)\rho\right) - Tr(\Pi_i(\mathcal{U}) \otimes \Pi_i(\mathcal{U})\rho) \quad (10)$$

$$P_2 = Tr(\Pi_i(\mathcal{U}) \otimes \Pi_i(\mathcal{U})\rho) \quad (11)$$

where $\Pi_i(\mathcal{U})$ is given in Eq. (7), and $\mathcal{U}$ is a unitary operation acting on a single-photon space. Integrating over many events, an intensity image is formed, determined by the distribution:

$$P(I_i = k) = \sum_{j=0}^{\lfloor k/2 \rfloor} P_1^{k-2j} P_2^j \quad (12)$$

Where $k$ is the photon count number at pixel $i$. Even if $P_1, P_2$ could be inferred, and the single-photon POVM in Eq. (7) was IC, dimensionality analysis shows the resulting measurement is not IC for the bi-photon space. Intuitively, this is expected, as the intensity measurement alone only provides a superposition of individual distributions and is blind to the correlations between photons. Thus, to recover the full multiphoton state, the measurement needs to be sensitive to multiphoton correlations. This insight leads to the following implications. Mathematically, the choice between the individual photons can be thought of as another DOF. As such, this could be achieved through a coupling system that inter-operates on the two photon subspaces using two-photon nonlinearities. However, realizing such nonlinearities at the single-photon level is experimentally challenging.

Instead, we can relax the requirement for a single intensity image while maintaining a single experimental setup. By replacing the standard camera with a detector capable of recording correlations, such as a Single-Photon Avalanche Diode (SPAD) array, one can perform spatial coincidence measurements [64–66]. Given a single-photon IC POVM $\Pi_i(\mathcal{U})$, the bi-photon POVM, $\Pi_i(\mathcal{U}) \otimes \Pi_j(\mathcal{U})$ is IC as well. This corresponds to a spatial coincidence measurement between the two photons after propagating them through the coupling system. ***This approach allows for full tomography of multiphoton hyperentangled states without any hardware reconfiguration between measurements.***

Physically, N-fold coincidence measurement scales exponentially with the number of photons, which poses a universal challenge of all multiphoton tomography, not unique to our method, but modern mitigation approaches can be applied [39]. Furthermore, certain schemes can access correlation functions using multiple intensity images only [65]. While such a shift would move the protocol from a single-image to a multi-image measurement, it would notably still avoid the need for active hardware reconfiguration.

Notably, this framework extends beyond photonics to other quantum processing platforms. For instance, recent work on ion trap quantum processors [41] demonstrated single-setting QST by projecting qubits into high-level energy states used as ancillas. Our framework captures this process by defining a qubit system ($d = 2, \mathrm{m} = 1$) with a coupler into a $D = 4$ space and utilizing a coincidence measurement as the measurement modality. This demonstrates that our framework, although intended for photonic systems, is more general and applicable to other quantum processing platforms that offer access to high dimensional states allowing for a minimal tomography scheme.

## 5. Discussion

To provide context for our approach, it is valuable to compare it with established scalable frameworks in QST. Modern frameworks such as compressed sensing tomography [30,31], adaptive tomography [32,33], threshold QST [34,35], and shadow tomography [36–38] have emerged as powerful tools to optimize the total amount of data required for state reconstruction. A critical distinction must be made between these methods and our proposed framework regarding the nature of the problem they address. Frameworks like the aforementioned are primarily designed to optimize the total number of measurements or the sample complexity required to characterize a state. In contrast, our approach specifically addresses the complexity of the physical measurement apparatus. By utilizing the inherent parallelism of intensity measurements, we eliminate the need for active hardware reconfiguration, such as rotating wave-plates or updating phase-masks, which are typically required for every term in the density matrix in traditional QST.

Additionally, we note that our method is highly complementary to these theoretical frameworks rather than a replacement for them. For example, combining our single-setup scheme with algorithms like compressed sensing could increase noise tolerance and reduce photon requirements. As a concrete example, if one has prior information about the state's rank, the optimization problem in Equation (9) can be updated to incorporate this information by minimizing the density matrix rank subjecting it to the constraints imposed by the measurements [40].

It is interesting to view the connection of our method to shadow tomography [36], and specifically classical shadows [37,38]. In the classical shadow framework, a quantum state is typically characterized by a collection of randomized measurements. In our approach, the intensity image itself serves as the classical shadow – a single, fixed classical representation that contains sufficient information to predict multiple properties of the state. The result of a linear function of the density matrix, $\langle O\rangle = Tr(O\rho)$, can be computed as a direct linear combination of the intensity image pixels, provided that the observable $O$ lies within the space spanned by the intensity POVM elements. Consequently, if only a specific subset of functions is required, the IC requirement of the system can be relaxed, allowing for a significant simpler coupler.

## 6. Conclusions

We introduced a general framework for the tomography of general single-photon states, entangled between multiple photonic DOFs, using a single intensity image obtained with a single experimental setup. Our framework hinges on the ability to construct a coupler that can mix between arbitrary spatial DOFs and non-spatial DOFs, enabling full encoding of the state's information into the spatial intensity pattern. This method eliminates the need for multiple measurement configurations, reducing experimental complexity and lowering errors from setup reconfiguration, and allowing for a faster reconstruction time. We demonstrated numerically how states entangled between their OAM and spin, and between their OAM and frequency, can be reconstructed using a coupler based on a MMF and the XPM effect. Importantly, we showed how these reconstruction schemes can be easily realized using the MMF classical TM and the introduced spatio-spectral TM allowing for a low computational cost. We also presented an extension of this method for reconstructing multiphoton hyperentangled states by replacing the single intensity with a single coincidence measurement setup. While this is more experimentally demanding, it enables full tomography of multiphoton quantum states in a single, static measurement setup.

We envision that this general approach can be applied across a wide range of platforms to enable the reconstruction of single-photon hyperentangled states and beyond. In particular, integrated photonic devices are well-suited to this framework, offering a path toward compact, scalable, and fully integrated solutions for quantum state characterization. We are currently working towards an experimental demonstration that showcase our method in the lab. Future directions include exploring spatial and time-bins coupling, and demonstrating the framework extension to multi-photon states.

## 7. Appendix A: Proof of Equation (7)

We prove that Eq. (8) is equivalent to Eq. (4). Substituting (7) into (8) and using the cyclic property of the trace, we find that

$$Tr\big(\rho\Pi_{\mathrm{i}}(\mathcal{U})\big) = Tr\left(\mathcal{U}\rho\mathcal{U}^{\dagger}(|r_{\perp};D\rangle\langle r_{\perp};D| \otimes I_m)\right) \tag{A1}$$

For any operator $\mathcal{O} \in \mathcal{H}_A$ and a system $\sigma \in \mathcal{H}_A \otimes \mathcal{H}_B$, the partial trace is exactly defined such that operating with $\mathcal{O}$ on $Tr_B(\sigma)$ is equivalent to operating on the entire system with $\mathcal{O} \otimes I_B$. Using $\mathcal{O} = |r_{\perp};D\rangle\langle r_{\perp};D|$ and $\sigma = \mathcal{U}\rho\mathcal{U}^{\dagger}$ gives us Eq. (4).

## 8. Appendix B: Numerical considerations of POVM description

Equation (4) gives a natural and simple description of the measurement process. A state $\rho \in \mathcal{H}_{spatial}^{d} \otimes \mathcal{H}_{non-spatial}^{m}$ propagates through the coupler $\mathcal{U} \in \mathbb{C}^{Dm\times dm}$, resulting in the output state $\mathcal{U}\rho\mathcal{U}^{\dagger} \in \mathcal{H}_{spatial}^{D} \otimes \mathcal{H}_{non-spatial}^{m}$. This state is then measured by a camera which is blind to the non-spatial DOF, hence the measurement is performed only on the spatial subsystem, mathematically described by the partial trace on the state. However, when solving the optimization problem (9), this description necessitates calculation of $\rho \rightarrow Tr_m(\mathcal{U}\rho\mathcal{U}^{\dagger})$ for every iteration, which is inefficient in time and memory complexity. Describing the measurement as a POVM (Eq. (8)) allows to calculate the POVM operators in Eq. (7) in advance, and during the optimization one only needs to perform the matrix multiplication between matrices of identical dimension $dm \times dm$.

Additionally, due to the unique structure of the POVM matrices in Eq. (7), also their calculation can be performed efficiently in terms of the required memory, which is necessary for large coupling matrices. Given three matrices, $A, C^T \in \mathbb{C}^{s\times dm}, B \in \mathbb{C}^{d\times d}$, where A is built by stacking $d$ blocks of dimension $s \times m$, and similarly $C$ is the concatenation of $d$ blocks of dimension $m \times s$. We mark these blocks $A_i^{s\times m}$ and $C_i^{m\times s}$. It can be shown that

$$A \cdot (B \otimes I_m) \cdot C = \sum_{i,j=1}^{d} B_{ij} \cdot A_i^{s \times m} C_j^{m \times s} \tag{B2}$$

Where $I_m$ is the $m \times m$ identity matrix and $B_{ij}$ is the $ij$ entry of $B$.

## 9. Appendix C: System Stability and Robustness to Characterization Errors

To evaluate the practical viability of our framework, we investigate the sensitivity of the reconstruction to noise encountered during the initial characterization of the system's response. In an experimental setting, the characterized unitary, $\tilde{\mathcal{U}}$, will deviate from the true physical $\mathcal{U}$, due to detector noise, or mechanical instabilities during calibration. We model this characterization error as independent, additive white Gaussian noise. To ensure a consistent comparison across different Hilbert space dimensions, we define the perturbed unitary as:

$$\tilde{\mathcal{U}} = proj\left(\mathcal{U} + \frac{\epsilon}{\sqrt{D}}\mathcal{G}\right)$$

Where $\mathcal{G}$ is a matrix of independent complex Gaussian random variable with zero mean and unit variance, and $D$ is the dimension of the unitary matrix. This normalization ensures that the norm of the perturbation remains proportional to $\epsilon$ regardless of the system size. To maintain physical consistency with the nature of the coupler, the matrix is subsequently projected back onto the space of unitary matrices.

The results of this sensitivity analysis are summarized in two key evaluations. First, we examine the impact of the perturbation strength $\epsilon$ on the experimentally measured TM used in Section 3.2 across varying system complexities. In Figure 5(a), we plot the reconstruction fidelity as a function of $\epsilon$ for different effective number of ancilla modes by selecting different mode groups. All test states remain four-dimensional, comprised of two OAM modes ($\pm l$) and two polarization states, where the OAM index determines the specific mode group the state is coupled to. The results demonstrate that robustness depends on both the number of ancillas and the intrinsic mixing properties within the mode group. Notably, the $6^{th}$ mode group ($l = \pm 5$) exhibits higher robustness to characterization error than the other groups tested. While higher-order groups possess more ancillas, this result suggest that the $6^{th}$ group provides higher intra-group crosstalk, facilitating more effective information spreading. Importantly, for most mode groups, the fidelity remains stable even under moderate characterization noise ($\epsilon < 0.1$).

Secondly, we explore the interplay between characterization noise and the signal-to-noise ratio (SNR) of the measured intensity image. Figure 5(b) presents a 2D map of the reconstruction fidelity as a function of the perturbation strength $\epsilon$ and the SNR for the $6^{th}$ mode group. As evidenced by the black contour line marking the 0.99 fidelity threshold, high-fidelity reconstruction is achievable across a wide operational regime. Even at a relatively low SNR of 15 dB, the system maintains near-perfect reconstruction provided the characterization error remains below $\epsilon \approx 0.1$.

All in all, this analysis confirms that the single-setup tomography framework described throughout this article is resilient to the typical experimental uncertainties encountered in MMF systems, and ensures high accuracy recovery of high dimensional hyperentangled quantum states even in the presence of errors in the characterization of the transformation of up to 10%.

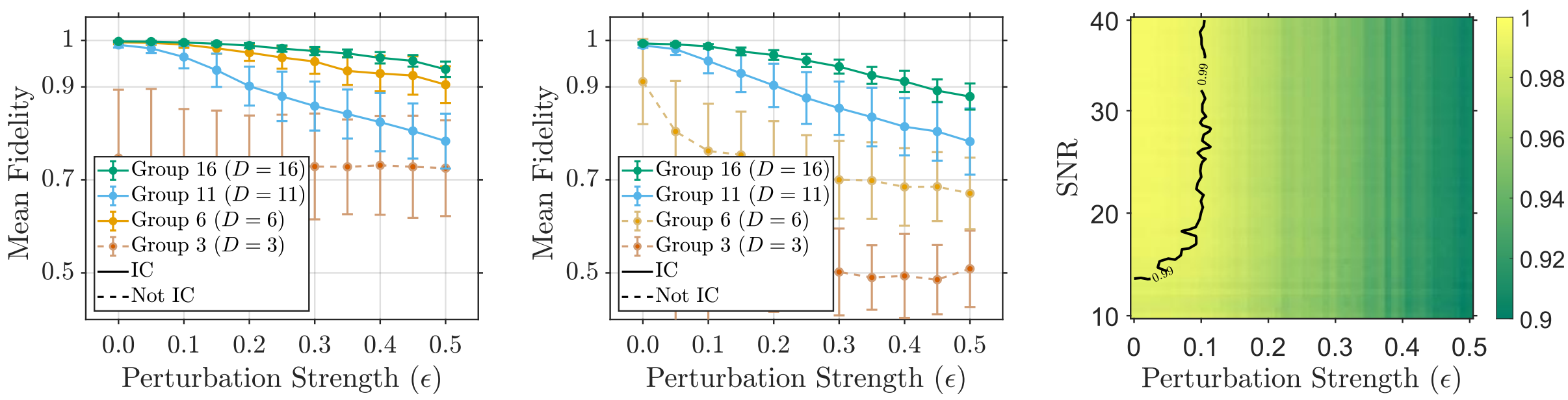

Figure 5: **Robustness of the reconstruction framework to characterization errors.** Mean fidelity as a function of perturbation strength $\epsilon$ for (a) 4-dimensional and (b) 8-dimensional hyperentangled states across different fiber mode groups. Dashed lines represent mode groups that do not satisfy the dimensionality condition ($D < dm$). Robustness is shown to depend on both the group size and the intrinsic mixing properties, with the $6^{th}$ mode group ($D = 6$) exhibiting high robustness. (c) 2D fidelity map as a function of perturbation strength and intensity image SNR for a 4-dimensional state coupled to the $6^{th}$ mode group. The black contour line highlights the regime where reconstruction fidelity exceeds 0.99, defining the operational stability requirements for the experimental setup.

## 10. Appendix D: OAM-Frequency bins states

Beyond spatial and spin DOFs, frequency-bin encoding provides a highly scalable resource for photonic quantum information processing. To extend our framework to the spectral domain, we leverage the optical Kerr effect - specifically, cross-phase modulation (XPM) between a structured pump pulse and the single-photon field [67,68].

The operational Hilbert space we consider consists of $m$ frequency bins - electromagnetic modes of identical spectral shape that are evenly spaced in frequency. By modulating the pump's temporal envelope at a frequency matching the bin spacing, we facilitate coupling between adjacent frequency bins, acting analogously to an electro-optic phase modulator used in frequency-bin photonic quantum information [69].

For a single-photon signal and a non-depleted pump, the interaction is governed by coupled nonlinear Schrodinger-like equations that remain effectively linear with respect to the signal field. This observation allows the entire complex interaction – spatial and frequency mixing, and temporal distortion, to be captured by a single spatio-spectral TM.

The spatio-spectral TM is built by characterizing the system's response to an orthogonal basis spanning all input spatial and frequency modes. To account for temporal broadening or dispersion at the output, we project the resulting fields onto a basis of orthogonal temporal modes, such as Hermite-Gaussian functions [70]. As with all other non-spatial DOFs, this temporal DOF is traced out when obtaining the intensity image.

We simulated [71] a 0.5 meter, graded index few-modes fiber. The system parameters were chosen to enhance interaction efficiency and computational feasibility. The pump pulse is $1nsec$ width, $0.5pJ$ at $1800nm$, supporting 3 spatial modes and modulated at $1THz$. At the single-photon wavelength of $900nm$, the fiber supports 10 spatial modes. These wavelengths are chosen both to reduce computational time (by reducing the number of modes at the pump's wavelength) and to ensure similar group velocities between the pump and single-photon, thereby enhancing interaction efficiency. While the pump and photon experience chromatic dispersion with opposite signs, this effect is negligible due to the short fiber length. The resulting spatio-spectral TM ($150x30$) maps the 30 input modes (10 spatial x 3 frequency) to an output space spanning 5 frequency bins and 3 temporal modes. The effective unitary $\mathcal{U}$ is then derived by transforming this TM into the OAM basis.

As shown in 6, the reconstruction fidelity follows the trends observed in the OAM-spin case, though with a higher sensitivity to the state's rank. At a photon count of $10^5$, the dependence on rank suggests that the 10-mode fiber with the frequency mixing doesn't provide enough redundancy. However, this is not a fundamental limit of the framework. In an experimental setting, using a standard MMF, increasing the number of ancilla modes can improve the mixing properties and information completeness of the measurement. Furth ermore, since MMFs can also support strong spatial-polarization coupling [55], this XPM-based method provides a path toward the simultaneous tomography of states entangled across spatial, spectral, and spin DOFs in a single measurement setup.

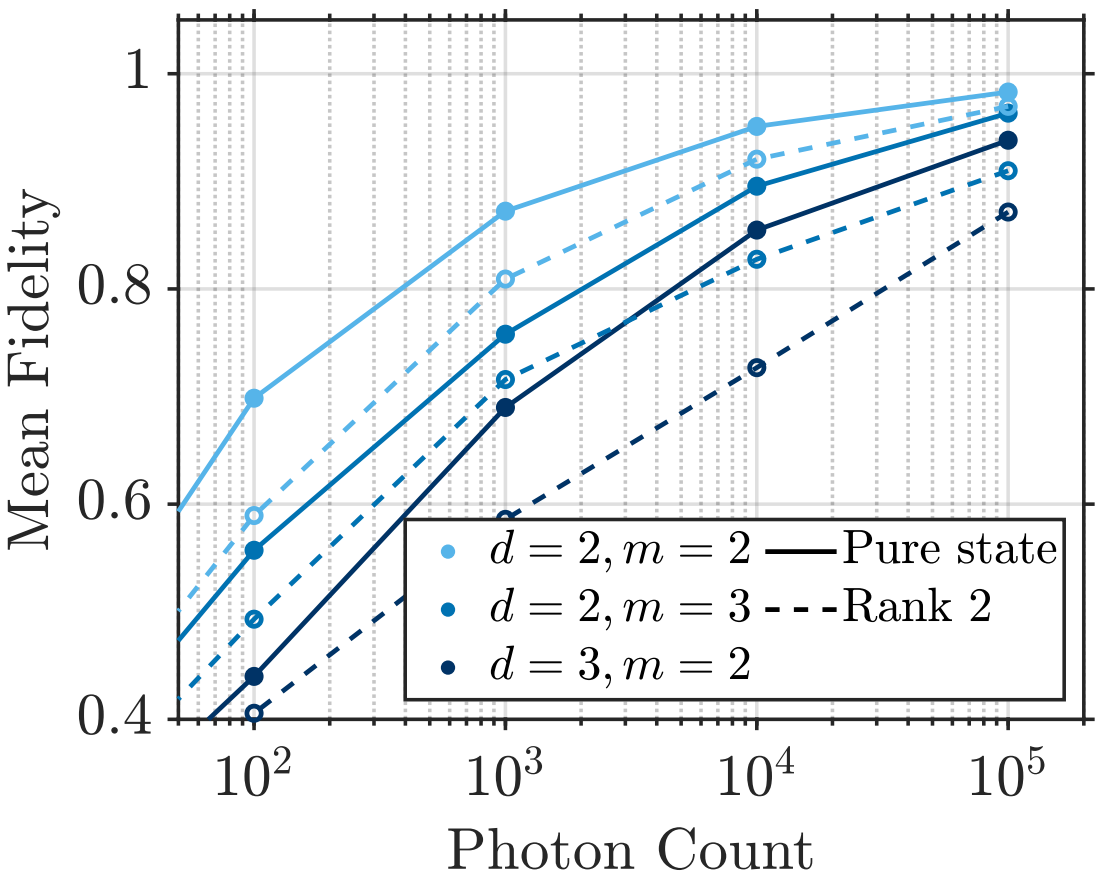

Figure 6: Reconstruction results of OAM-Frequency states plotted for different dimensions and ranks of the initial state.

**Disclosure.** The authors declare no conflicts of interest.

**Funding.** This project was funded by the German-Israeli Program DIP.

**Data availability.** All data generated and analyzed during this study are available from the corresponding author on reasonable request.